\documentclass[12pt]{iopart}
\usepackage{graphicx}
\usepackage{iopams}

\begin{document}

\title[Ultra-compact radio sources and the isotropy and homogeneity of the Universe]{Ultra-compact
radio sources and the isotropy and homogeneity of the Universe}
\author[J. C. Jackson]{J. C. Jackson$^{1}$\thanks{}\\
$^{1}$Division of Mathematics, School of Computing, Engineering and Information Sciences,
Northumbria University, Newcastle NE1 8ST, UK\\
}





\begin{abstract}
A 2.29 GHz VLBI all-sky survey of ultra-compact radio sources has formed the basis of a
number of cosmological investigations, which examine the relationship between angular-size
and redshift.  Here I use a sample of 468 such sources with $0.5< z\leq 3.787$, to
investigate the isotropy of the Universe.  The sample is divided into hemispherical
sub-samples, over an all-sky $5^\circ\times 5^\circ$ array, each of which is allowed to
determine a value of $\Omega_{\mathrm{m}}$, assuming that we are living in a spatially 
flat homogeneous isotropic $\Lambda$CDM model.  If we regard the latter as a null 
hypothesis, then it fails the test -- the results show significant anisotropy,
the smallest value of $\Omega_{\mathrm{m}}$ being towards $(l,b)=(253.9,\,24.1)^\circ$,
the largest in the opposite direction.  This is close to the CMB dipole axis, but in
the obverse sense.  This is interpreted as meaning that the Universe is not spatially
homogeneous on the largest scales, and is better represented at late times by a spherically
symmetric model with a density enhancement at its centre. 

\bigbreak
\end{abstract}

\section{Introduction}

In the large the Universe around us exhibits a remarkable degree of spherical
symmetry, the most stringent constraint on anisotropy being provided by observations
of the Cosmic Microwave Background (CMB).  The near isotropy of the latter was 
apparent at the time of its discovery (Penzias \& Wilson 1965; Wilson \& Penzias 1967).
Early COBE results showed a dipole component of $\Delta T=3.335$ mK ($0.12$ per cent),
and a monopole temperature of $2.725$ K (Kogut et al. 1993; Mather et al. 1999).
After removing the dipole component the rms sky variation is 
$\Delta T=45.3~\mu$K (0.0017 per cent) (see for example Percival et al. 2002)
over the multipole range $2\leq l \leq 1500$.  If we accept the Copernican principle
then these observations mean that the Universe is homogeneous and isotropic; the
corresponding Friedmann--Lema\^\i tre--Robertson--Walker (FLRW) framework has been
the the basis of most studies in observational cosmology.

\vskip 0.6cm
\nobreak
\noindent
\ddag E-mail: john.jackson@northumbria.ac.uk

However, while the Copernican principle remains untested, inhomogeneous models
should not be dismissed (Clarkson \& Maartens 2010; Ellis 2011).  I report here
a test of isotropy based upon the angular-size/redshift relationship,
using ultra-compact radio sources as standard measuring rods; these objects have
angular diameters in the milliarcsecond (mas) range, and linear sizes of order several parsecs.
In fact the test reveals significant anisotropy, a tentative interpretation of which is 
that the Universe is not spatially homogeneous on the largest scales, and is better represented 
at late times by a spherically symmetric model with a density enhancement at its centre.  
Antoniou \& Perivolaropoulos (2010) have already looked at Union2 SnIa dataset in this context, which shows a similar anisotropy; my approach closely follows theirs.  {More recently, Longo (2012)
has reported an anomaly in the angular distribution of quasar magnitudes, which is interpreted as
evidence for a large-scale distant inhomogeneity.}

\section{Ultra-compact radio sources}

The data to be used derive from an ancient VLBI survey of such sources at 2.29 GHz
(Preston et al. 1985, hereafter referred to as P85).  This survey employed a world-wide
array of dishes, forming an interferometric system with an effective baseline of about
$8\times 10^7$ wavelengths; the survey gave total and correlated flux densities
(fringe amplitudes) for 917 objects, $S_{\mathrm{t}}$ and $S_{\mathrm{c}}$ respectively.
The ratio $\Gamma=S_{\mathrm{c}}/S_{\mathrm{t}}$ is a measure of fringe visibility,
from which angular size can be estimated (Thompson, Moran \& Swenson 1986; Gurvits 1994):

\begin{equation}\label{A}
\theta={2\sqrt{-\ln \Gamma \ln 2} \over \pi B},
\end{equation}

\noindent
where $B$ is the interferometer baseline, in wavelengths.

\begin{figure}[here]
\vspace{-4.5cm}
\begin{center}
\includegraphics[width=14.5cm]{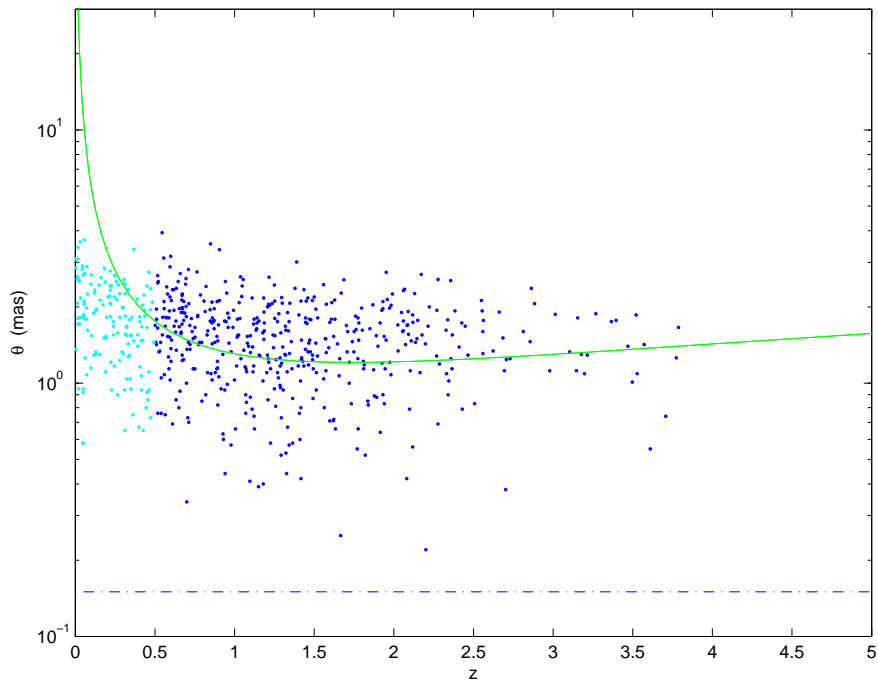}
\end{center}
\vspace{-5.0cm}
\caption{Angular-diameter/redshift diagram for the full sample of 468 sources { (blue points)},
and the corresponding best-fitting curve $\Omega_{\mathrm{m}}=0.29$ and $d=7.76h^{-1}$ pc.
The blue dashdot line is an estimate of the resolution limit $\theta_{\mathrm{lim}}=0.15$ mas.
{ The 145 cyan points have ${z\leq 0.5}$, and are not used in this investigation.}}
\label{FigA}
\end{figure}

The potential of P85 in this context was first noted by Gurvits (1994),
who considered a sample comprising 258 objects with redshifts $z>0.5$.
Using the same data set, Jackson \& Dodgson (1997) extended this work to the full
$\Omega_{\mathrm{m}}$/$\Omega_\Lambda$ plane, finding best values $\Omega_{\mathrm{m}}=0.2$
and $\Omega_\Lambda=0.8$, if the Universe is spatially flat, later refined to
$\Omega_{\mathrm{m}}=0.24$, $\Omega_\Lambda=0.76$ (Jackson 2004).  The sample was updated
with regard to redshift (Jackson \& Jannetta 2006), to give 613 objects with $0.0035\leq z\leq 3.787$,
of which 468 have $z>0.5$, which sub-set is used in this investigation.  There are
several reasons for ignoring sources with $z<0.5$.  As z falls below 0.5,
the epoch of quasar formation comes to an end, and the nature of the population
changes dramatically { (see the cyan points in Fig. \ref{FigA})};
there is a correlation between linear size and radio luminosity -- the weaker sources
are distinctly smaller, which would introduce an unacceptable selection bias.
There is no evidence of such an effect when $z>0.5$ (Jackson 2004); if it is there
it is not pronounced.  A residual bias should not appear to be anisotropic.

A model of these objects, which gives an account of their status 
as standard measuring rods, is discussed in Jackson (2004).
The model is supported by VLBA images (Kovalev et al. 2005),
which show a bright compact `core' at the end of a one-sided jet; the `core' is
believed to be the base of a continuous jet, rather than the nucleus
of the object.  Beamed emission from the compact core dominates the observed
structure.  The importance of D\H oppler beaming cannot be over-stated; the rest-frame
linear size of mas source components is known to be an increasing function of
wavelength (Marscher \& Shaffer 1980; Pearson \& Readhead 1981);
at first sight this would mean that the effective linear size should actually be a
decreasing function of $z$.  However, as $z$ increases a larger D\H oppler boost 
factor ${\cal D}$ is required; it turns out that the ratio ${\cal D}/(1+z)$
is approximately fixed (Jackson 2004), 
so that the emitted frequency ${\cal D}\nu_\mathrm{r}/(1+z)$ also fixed,
where $\nu_\mathrm{r}$ is the fixed reception frequency.  Note that this ratio is
not necessarily unity, but the fact that it is fixed means that the interferometric
angular sizes upon which this work is based are fit for purpose.  This is
another reason for ignoring sources with $z<0.5$, which appear to be
non-relativistic.  Again a residual bias should not appear to be anisotropic.

A second point which needs clarification is resolution.  With a baseline
$B=8\times 10^7$, the Rayleigh resolution limit is about 2.6 mas. The mean angular
size of the sources in question is 1.52 mas, somewhat below the Rayleigh limit.
However, it is well-known that a simple interferometric technique can achieve
a degree of super-resolution; the matter is discussed in detail in Kovalev et al. (2005). 
The limiting factor is signal--to--noise ratio $SNR$; for a simple Gaussian source
the limiting angular size is

\begin{equation}\label{B}
\theta_{\mathrm{lim}}
=b\left[{4\ln 2 \over \pi}\,\ln\left({SNR \over SNR-1}\right)\right]
^{1/2},
\end{equation}

\noindent
where $b$ is the half-power beam width, which I have taken to be be half the fringe spacing, 
thus $b=(2B)^{-1}$.  As to the appropriate $SNR$, two figures are mentioned in Preston et al.
(1985).  The first is described as a systematic error of about 10 per cent; 
a systematic percentage error should not affect measures of fringe visibility,
but I note that the corresponding resolution is $\theta_{\mathrm{lim}}=0.20$ mas.
The second figure is an absolute random error of 0.2 Jy in the correlated flux density,
in which case the $SNR$ would vary from source to source; the median SNR value for those
sources used in this invstigation is 17, giving $\theta_{\mathrm{lim}}=0.15$ mas.
These figures are indicative, but do confirm that lack of adequate angular resolution
is unlikely to be an issue here.
(However, for a radically different point--of--view see Pashchenko \& Vitrishchak 2011).

\begin{figure}[here]
\vspace{-4.5cm}
\begin{center}
\includegraphics[width=14.5cm]{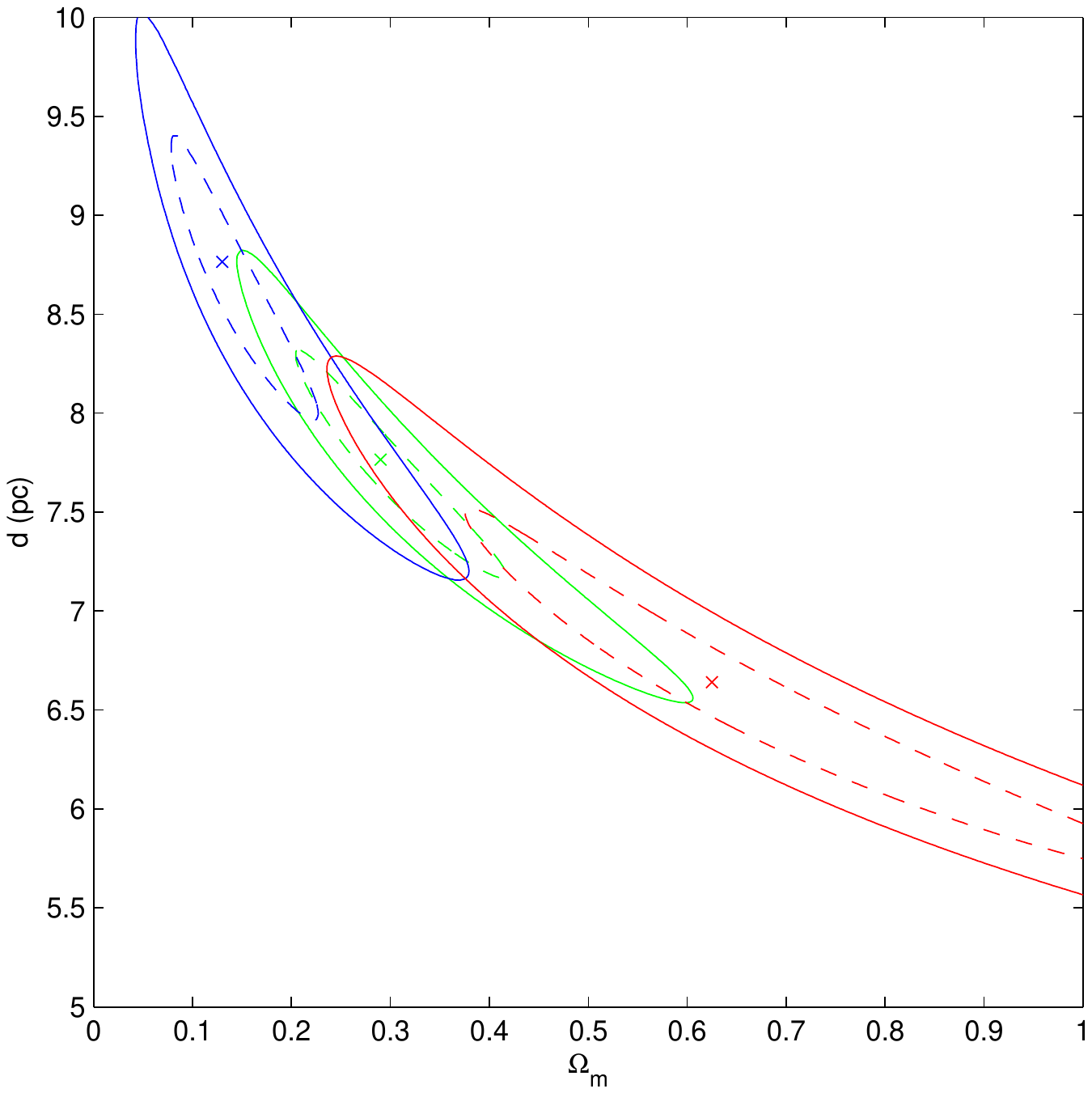}
\end{center}
\vspace{-5.0cm}
\caption{Confidence regions, one and two $\sigma$: green corresponds to the the full
sample, blue and red to opposing extreme hemispheres. These are marginalized, so that 
projection on to each axis gives confidence intervals for the respective parameter.}
\label{FigB}
\end{figure}

Fig. \ref{FigA} is the angular-diameter/redshift diagram for the full sample of 468 sources
discussed above.  The continuous green line corresponds to the best-fitting spatially
flat homogeneous isotropic $\Lambda$CDM model, characterized by two free parameters,
the matter density $\Omega_{\mathrm{m}}$ and the intrinsic linear size $d$.  The best values
are $\Omega_{\mathrm{m}}=0.29$ and $d=7.76h^{-1}$ pc ($H_0=100h$ km sec$^{-1}$ Mpc$^{-1}$),
being a fit to the unadorned unbinned data points; Fig. \ref {FigB} shows the corresponding
confidence regions (the green curves).   

I have undertaken a similar analysis of the Caltech-Jodrell Bank 5 GHz survey,
giving similar results, which survey cannot however be used in this context because
it is not an all-sky one (Taylor et al. 1996; Jackson 2008).

\section{Anisotropy}

\begin{figure}[here]
\vspace{-4.5cm}
\begin{center}
\includegraphics[width=14.5cm]{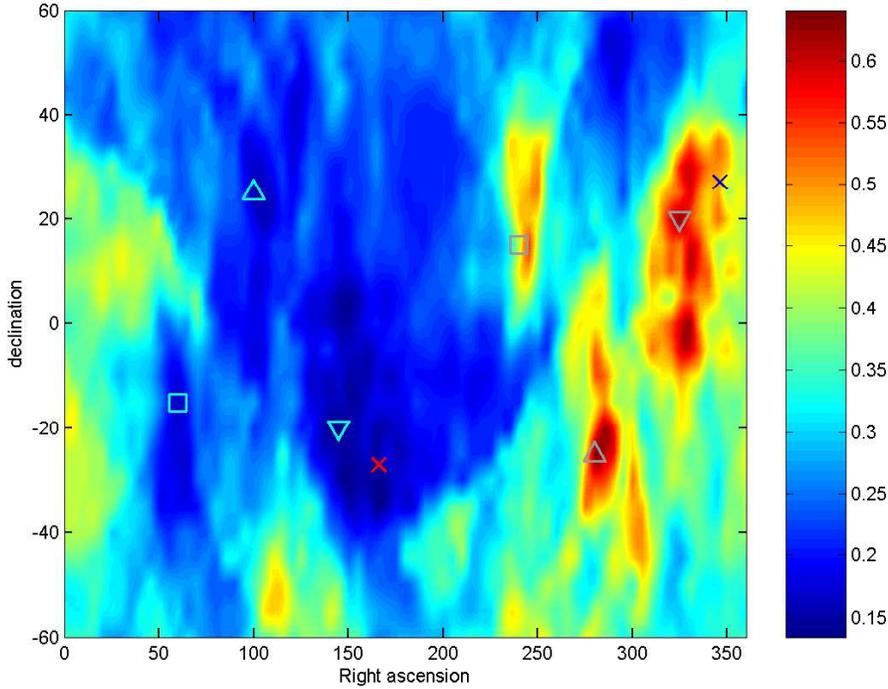}
\end{center}
\vspace{-5.0cm}
\caption{Pseudocolour plot showing the distribution of $\Omega_{\mathrm{m}}$ over the sky;
full sample of 468 sources, $0.5<z\leq 3.787$.  There are no prominent features in the
polar regions. Crosses indicate the CMB dipole in the Local Group frame; the
other symbols delineate prominent features for further reference.}
\label{FigC}
\end{figure}

To test isotropy, I have considered hemispheres, typically containing $234\pm 11$ sources,
which data points are allowed to determine $\Omega_{\mathrm{m}}$ and $d$;
the values so determined now depend upon the particular hemisphere, defined by the 
Right ascension and declination ${(\alpha,\delta)}$ of its axis.  { The coordinates
used here are B1950, as listed in P85.}  
Fig. \ref {FigC} is a pseudocolor Mercator projection based upon an evaluation of
$\Omega_{\mathrm{m}}$ on a $5^\circ\times 5^\circ$ grid over the range $0^\circ\leq \alpha <360^\circ$,
$-60^\circ\leq \delta \leq 60^\circ$.  The plot shows a pronounced asymmetry,
a measure of which is the parameter { ${D=\Delta \Omega_\mathrm{m}}$ evaluated over
opposing hemispheres}.  The global maximum value is $D_{\mathrm{max}}=0.626-0.135=0.491$ 
towards $(l,b)=(253.9,\,24.1)^\circ$, which corresponds to the smaller value of
$\Omega_{\mathrm{m}}$, indicated by the cyan $\nabla$ mark in Fig. \ref {FigC};
the opposite direction is indicated by the grey $\nabla$ mark.  

The red and blue $\times$ marks indicate the CMB
dipole in the Local Group frame, with velocity $627\pm 22$ km sec$^{-1}$,
the `hot' direction being $(l,b)=(276\pm 3,\,30\pm 3)^\circ$ (Kogut et al. 1993).
The proximity of these two directions is quite striking; however, the CMB
dipole cannot be attributed to a peculiar velocity induced by gravitational attraction
towards the region of enhanced density, which would be in the wrong direction,
a point to which I shall return later.  The features indicated by the $\triangle$
and square marks are for later reference.  The figure reported by Antoniou \& Perivolaropoulos (2010)
is $D_{\mathrm{max}}=0.30-0.19=0.11$ towards $(l,b) = (309^{+23}_{-03},\,18^{+10}_{-11})^\circ$,
which again corresponds to the smaller value of $\Omega_{\mathrm{m}}$.
The Union2 SnIa dataset covers the redshift range $0.015\leq z\leq 1.4$ (Amanullah et al. 2010). 
 
The blue and red confidence regions in Fig. \ref {FigB} correspond to the hemispherical samples
which determine $(\Omega_{\mathrm{m}})_{\mathrm{min}}$ and $(\Omega_{\mathrm{m}})_{\mathrm{max}}$,
which samples I shall call $\hbox{S}_{\mathrm{min}}$ and $\hbox{S}_{\mathrm{max}}$.
I have considered one way in which the differences shown in Fig. \ref {FigB} might
be spurious.  The fitting problem is close to degeneracy with respect to the
parameters $\Omega_{\mathrm{m}}$ and $d$, because we cannot use nearby sources,
which would otherwise fix the latter (Jackson \& Dodgson 1996).
I have considered the possibility that there are in fact no significant differences between
$\hbox{S}_{\mathrm{min}}$ and $\hbox{S}_{\mathrm{max}}$, other than small accidental
ones which due to the degeneracy are nevertheless large enough to bring about
the divergences seen in Fig. \ref {FigB}.  This possibility is discounted by the
following considerations; I have further divided the two hemispherical samples,
into $\hbox{S}_{\mathrm{min}}(z\!\leq\! 1.5)$ and $\hbox{S}_{\mathrm{min}}(z\!>\!1.5)$,
and similarly for $\hbox{S}_{\mathrm{max}}$.  It is indeed the case that the low-redshift
sub-samples are virtually identical, but in the high-redshift case there is a
statistically significant difference: the two mean angular-sizes are
${\overline {\theta}}_{\mathrm{min}}(z\!>\!1.5)=1.35\pm 0.05$ mas and
${\overline {\theta}}_{\mathrm{max}}(z\!>\!1.5)=1.56\pm 0.06$ mas ($1~\sigma$ errors),
which difference is not accounted for by a difference in the respective
values of $\overline{z}$. The two values of ${\overline {\theta}}(z\!>\!1.5)$
are the driving force behind the results presented here.   

\begin{figure}[here]
\vspace{-4.5cm}
\begin{center}
\includegraphics[width=14.5cm]{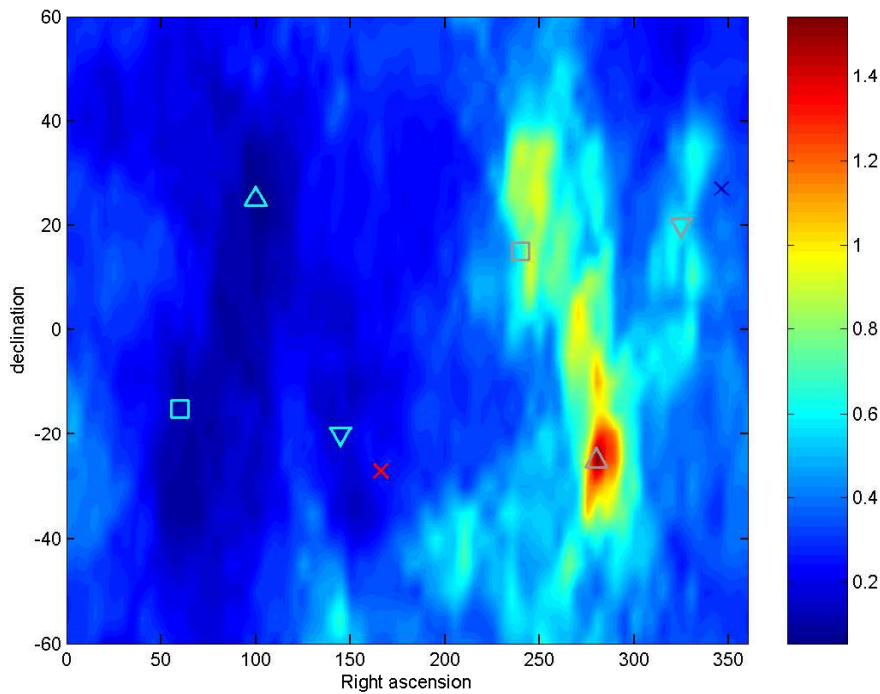}
\end{center}
\vspace{-5.0cm}
\caption{Pseudocolour plot showing the distribution of $\Omega_{\mathrm{m}}$ over the sky;
restricted sub-sample of 380 sources, $0.5<z\leq 2.0$. The features highlighted in
Fig. \ref {FigC} have not been eliminated by dropping the high-redshift sources.
}
\label{FigD}
\end{figure}

I have considered sensitivity to the exclusion of high-redshift sources,
by looking at restricted samples defined by $0.5<z\leq z_{\mathrm{max}}$.
The pattern remains reasonably stable as long as ${z_{\mathrm{max}}}~_\sim^>~2.0$
(see Fig. \ref {FigD}); at $z_{\mathrm{max}}=1.5$ there is little trace of the
original structure.  The results presented here thus do not depend upon a
few high-redshift objects.

\begin{figure}[here]
\vspace{-4.5cm}
\begin{center}
\includegraphics[width=14.5cm]{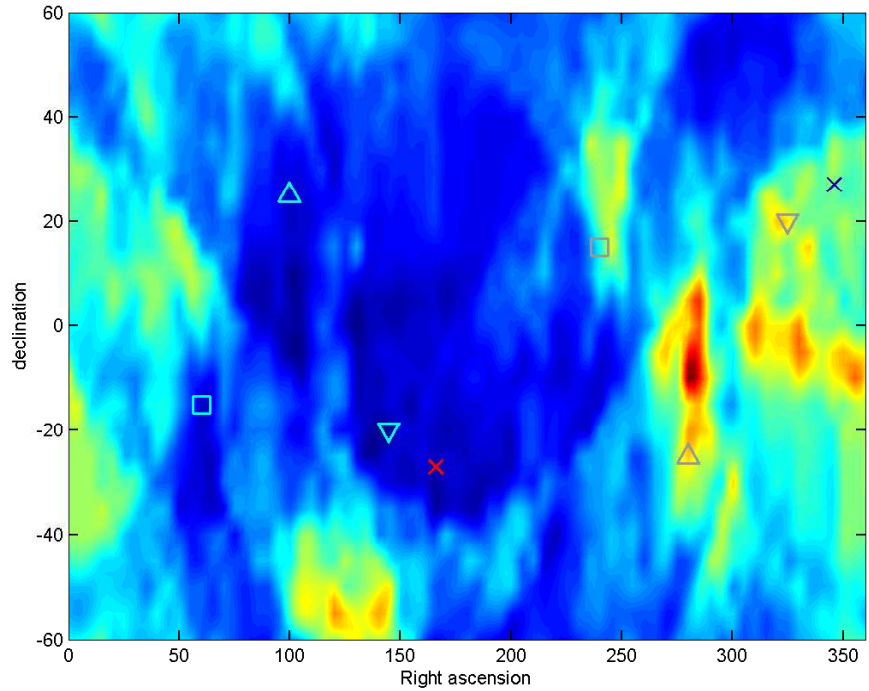}
\end{center}
\vspace{-5.0cm}
\caption{Pseudocolour plot showing the distribution of $\Omega_{\mathrm{m}}$ over the sky;
even sub-sample of 234 sources, see text.}
\label{FigE}
\end{figure}

\begin{figure}[here]
\vspace{-4.5cm}
\begin{center}
\includegraphics[width=14.5cm]{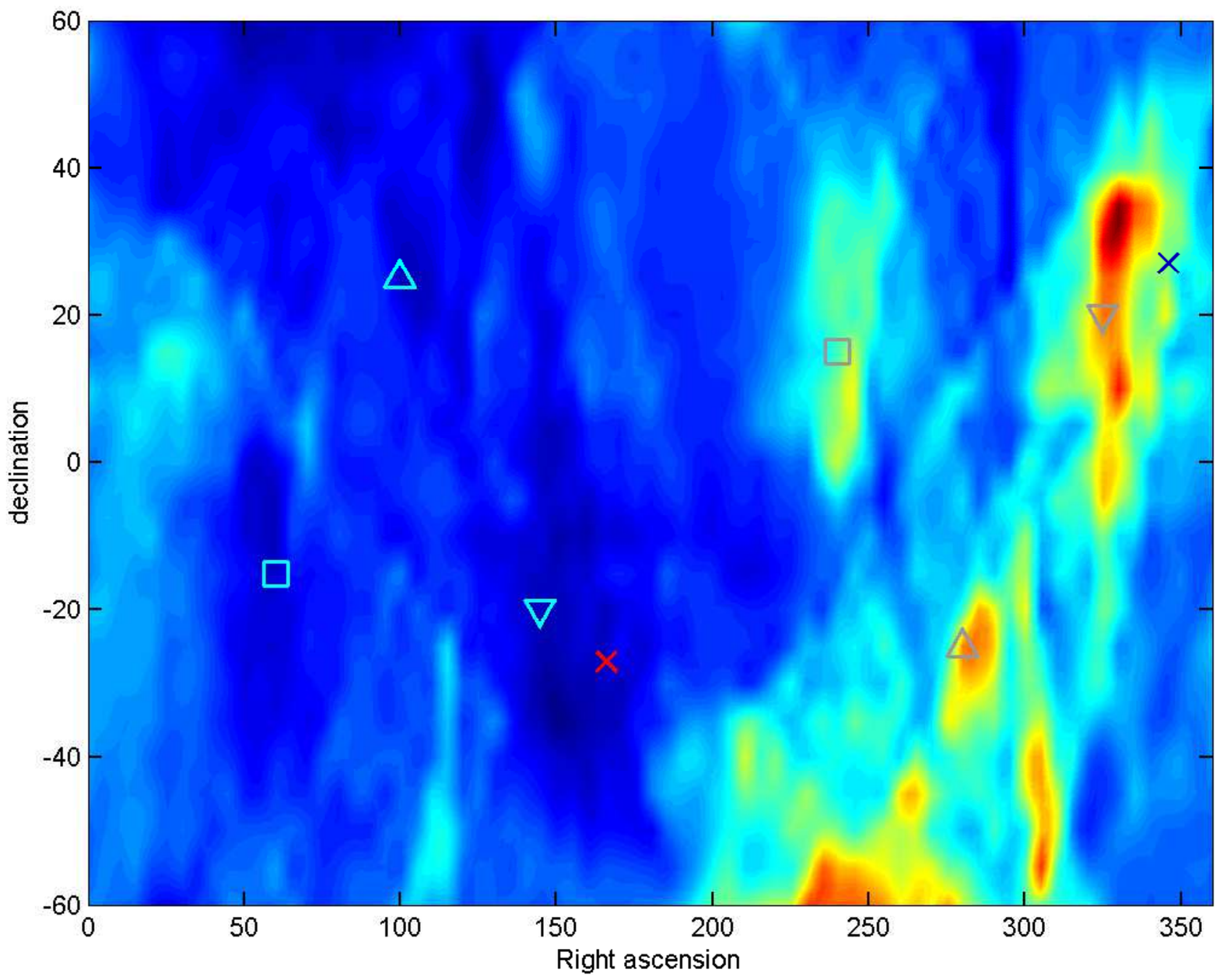}
\end{center}
\vspace{-5.0cm}
\caption{Pseudocolour plot showing the distribution of $\Omega_{\mathrm{m}}$ over the sky;
odd sub-sample of 234 sources, see text.}
\label{FigF}
\end{figure}

Could statistical fluctuations in the distribution of for example high-redshift
sources between opposing hemisperes account for the apparent asymmetry?
I have evaluated $D_{\mathrm{max}}$ over an ensemble of mock samples, 
generated by randomizing source positions while leaving the corresponding redshift
and angular-size data unchanged.  The distribution of $D_{\mathrm{max}}$ values so
produced is broad, with a median value of 0.508.  The above value $D_{\mathrm{max}}=0.491$
is thus { inconclusive}.  As an alternative test, I have first subjected
the full list of 468 sources to a random shuffle, and then divided the list into 
two independent sub-samples, the first comprising all even-numbered members, the
second all odd-numbered such members; Figs. \ref {FigE} and \ref {FigF} show the
hemispherical distribution of $\Omega_{\mathrm{m}}$ over the sky for the even and odd
sub-samples respectively.  If the prominent features delineated in Fig. \ref {FigC}
are due entirely to statistical fluctuations, then Figs. \ref {FigE} and \ref {FigF}
would not be expected to show the same features, whereas it is quite clear that they do.
Figures \ref {FigE} and \ref {FigF} are very similar; they are highly correlated,
with a correlation coefficient of 0.413.  Remembering that the even and odd
sub-samples have no members in common, the probability of this situation arising by
chance is less than $1\times 10^{-9}$ (see for example Freund \& Walpole 1980).
I conclude that the observed asymmetry is intrinsic.  There are $468$! distinct
permutations of the original list, and I have examined a modest number of these;
Figs. \ref {FigE} and \ref {FigF} represent a typical example.

\begin{figure}[here]
\vspace{-4.5cm}
\begin{center}
\includegraphics[width=14.5cm]{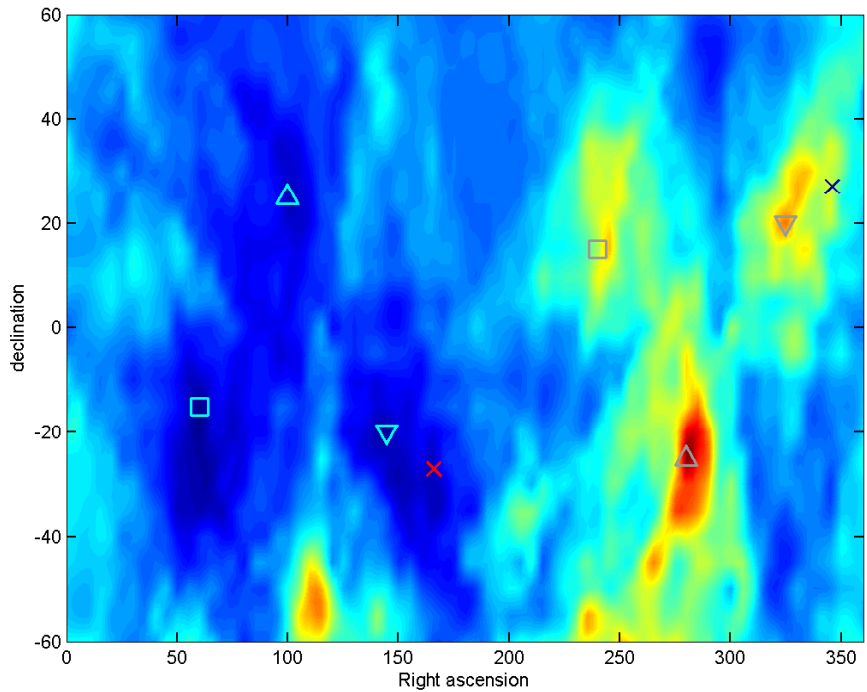}
\end{center}
\vspace{-5.0cm}
\caption{Pseudocolour plot showing the distribution of $\Omega_{\mathrm{m}}$ over the sky;
restricted sample comprising 339 pure P85 sources, $0.5<z\leq 3.787$.}
\label{FigG}
\end{figure}

{
The remaining possibility is that the apparent anisotropy is an instrumental effect, 
affecting all of the above-mentioned samples and sub-samples in like fashion.
In this respect a suspect feature of the full sample of 468 sources is that 
not all of the flux measures derive from the same system; 339 are pure P85,
in that they have both correlated and integrated flux densities listed in P85;
the remaining 129 have correlated fluxes in P85, without the corresponding
integrated fluxes; in these cases the latter were taken from the Parkes catalogue
PKSCAT90 (Wright \& Otrupek 1990).  As the correlated fluxes are the critical ones, I do not
regard this as a serious deficiency; to confirm this I have discarded the 
`mixed' sources and repeated the primary computation described above:
Fig. \ref {FigG} shows the ${\Omega_{\mathrm{m}}}$ distribution determined by the
set of pure P85 sources, which is very similar to Fig. \ref {FigC}.  The anisotropy
is clearly not an artefact of instrumental inhomogeneity.  Finally there is the
matter of uniformity of sky cover.  Without the additional Parkes sources the pure
P85 sample is deficient in southern-hemisphere ones (${\delta>-41.9^\circ}$);
with the additional Parkes sources the sample is uniform, the mean number per
hemisphere being ${233.95\pm 11.16}$, very close to the binomial figure ${234\pm 10.82}$.
}

\section{Interpretation and a toy model}

We cannot of course maintain that the Universe is well-represented by a flat homogeneous
isotropic $\Lambda$CDM model, but that a different such model is required according to
direction.  Nevertheless, I believe that these results mean that there is more dark
matter in some directions than in others, whatever the geometrical setting.  To see this,
we consider a source represented by a small plane circular disc, perpendicular to the 
light-ray going from its centre to the observer.  In vacuum, the bundle of rays 
going from the edge of the disc to the observer defines a cone with the observer
at its apex; in the non-vacuum case the cone is refocussed by the gravitational attraction
of the matter within it, as we trace the rays back towards the source, which refocussing
has a magnifying effect.  Indeed if the source is sufficiently distant then the cone
begins to reconverge, and the apparent size of the source begins to increase
(see for example Ellis \& Tivon 1985).  A reasonable explanation of the fact that the sources in the
hemispherical sample $\hbox{S}_{\mathrm{max}}$ appear to be systematically larger than those
in $\hbox{S}_{\mathrm{min}}$, is that the rays in the direction of the former are passing
through denser matter than those in the direction of the latter.  In one or two simple
cases an exact analytical treatment is possible (Jackson \& Dodgson 1996), based upon the
Ehlers-Sachs equation (Ehlers \& Sachs 1959, Sachs 1961, Pirani 1965), which rederives some
of the elementary cosmological results relating to distance.  

The matter concentration revealed here cannot be behaving like a simple attractor,
because the motion of the Local Group would be towards the latter, whereas it appears
to be in the opposite direction.  However, this is exactly what an off-centre 
fundamental observer (at rest relative to the local Hubble flow) would expect
to see in a spherically-symmetric dust-filled inhomogeneous model with a central
concentration (Raine \& Thomas 1981; Maartens et al. 1996a; Maartens et al. 1996b;
Humpreys, Maartens \& Matravers 1997).
According to this picture the apparent motion of the Local Group is due largely to a peculiar cosmological redshift, rather than a peculiar flow (Paczynski \& Piran 1990).  

Large-scale coherent peculiar motions would be a feature of such a model, which 
must be compared with observations.  The local such flow is believed to be 
due to the Great Attractor, a massive concentration of elliptical galaxies
apparently at rest in the CMB frame, at a distance of $43.5h^{-1}$ Mpc towards
$(l,b)=(307,\,9)^\circ$ (Lynden-Bell et al. 1988); this result was based upon a sample
of 400 elliptical galaxies closer than $80h^{-1}$ Mpc.  Infall (rather than uniform
bulk flow) gives a good account of the near-side flow; however, evidence for infall
from the far side of the attractor is inconclusive (Dressler \& Faber 1990).
Watkins, Feldman \& Hudson (2009) and Feldman, Watkins \& Hudson (2010) examined
a compilation of similar samples,
and concluded that the mean peculiar motion of galaxies within a spherical
volume of radius c. $100h^{-1}$ Mpc is $416\pm 78$ km s$^{-1}$, towards
$(l,b)=(282\pm 11,\,6\pm 6)^\circ$, and that this velocity appears to be increasing
with distance. As the authors remark, a flow of this amplitude on such a large
scale is not expected in the WMAP5-normalized $\Lambda$CDM cosmology.
  
Kashlinsky et al. (2008, 2009, 2010) have looked for CMB temperature fluctuations
induced by the kinematic Sunyaev-Zel'dovich effect, from which the velocities of
the associated clusters can be deduced.  These authors report a positive
detection with bulk velocities in the range $600$ to $1000$ km sec$^{-1}$ towards
$(l,b)=(283\pm 14,\,11\pm 14)^\circ$, with most of the signal arising from a shell
between $150h^{-1}$ and $600h^{-1}$ Mpc; { such flows are not easily accomodated
within the standard inflationary $\Lambda$CDM model (Atrio-Barandela et al. 2010).
However, the latter results have been discounted by others
(Keisler 2009; Osborne et al. 2011; Mody \& Hajian 2012; Hand et al. 2012).

For purposes of comparison, I have considered a simple spherically symmetric
zero-energy Newtonian model:

\begin{equation}\label{C}
v^2={2GM \over r(M)}+{1 \over 3}\Lambda r(M)^2,
\end{equation}

\noindent
where $r(M)$ is the comoving radius of a sphere of mass $M$ centred on the point
of maximum concentration, and $v(r)$ is the radial velocity relative to the centre.
Introducing $\bar\rho$ as the mean density within a sphere of radius $r$,
equation (\ref{C}) becomes 

\begin{equation}\label{D}
v^2={8\pi G\bar\rho r^2 \over 3}+{1 \over 3}\Lambda r^2.
\end{equation}

\noindent
We consider a snapshot and radial power series expansion about a point at $r_0$:

\begin{equation}\label{E}
v=v_0+(r-r_0)\left({\mathrm{d}v \over \mathrm{d}r}\right)_0
+{1 \over 2}(r-r_0)^2\left({\mathrm{d}^2v \over \mathrm{d}r^2}\right)_0+......
\end{equation}

\noindent
The first-derivative in equation (\ref{E}) gives the local Hubble flow:

\begin{equation}\label{F}
v_{_\mathrm{H}}=H_0\Delta r=\left[-{4\pi G}A^{-1/2}(\bar\rho-\rho)+A^{1/2}\right]_0\Delta r,
\end{equation}

\noindent
where $\Delta r=r-r_0$ and $A=8\pi G\bar\rho/3+\Lambda/3$. The second derivative gives a
local radial peculiar flow; on a shell of radius $\Delta r$, this is (dropping the subscript):
\medbreak

\begin{equation}\label{G}
v_{_\mathrm{B}}=
\left\{-{A^{-1} \over r}\left[4\pi G(\bar\rho-\rho)\right]^2
\!+A^{-1/2}4\pi G\!\left[{2 \over r}(\bar\rho-\rho)+\!{\mathrm{d}\rho \over \mathrm{d}r}\right]\right\}
\!\!{{\Delta r}^2 \over 2}.
\end{equation}

\noindent
A positive value for $v_{_\mathrm{B}}$ corresponds to peculiar motion away fron the centre.
We are considering distributions in which $\bar\rho-\rho>0$ and $\mathrm{d}\rho/\mathrm{d}r<0$,
so that either sign is possible; for example, if $\bar\rho-\rho>>|\mathrm{d}\rho/\mathrm{d}r|$ and
$\bar\rho$ and $\rho$ are not too different, the dominant term is

\begin{equation}\label{H}
v_{_\mathrm{B}}
\!\sim A^{-1/2}4\pi G\rho\left({\bar\rho-\rho \over \rho}\right)\!{\Delta r^2 \over r}
\sim {1 \over 2}\left({\bar\rho-\rho \over \rho}\right)\!{\Delta r \over r}H_0\,\Delta r.
\end{equation}

\noindent
Taking $(\bar\rho-\rho)/\rho$ and $\Delta r/r$ to be $0.20$, $\Delta r=100$ Mpc and
$H_0=70$ km sec$^{-1}$ Mpc$^{-1}$, equation (\ref{H}) gives $v_{_\mathrm{B}}=140$ km sec$^{-1}$.
The main point of this rough estimate is to show that the model is not obviously at variance
with the developing observational evidence, and that equations (\ref{F}) and (\ref{G})
are sufficiently flexible to accomodate possible contingencies.

\section{Acknowledgments}
It is a pleasure to thank Dr. William Stoeger of the Vatican Observatory
Research Group, Steward Observatory, University of Arizona, who encouraged me 
to undertake this extension of the earlier work of myself and my colleagues;
also to thank Richard Jackson for computational assistance.

\section{References}

{\obeylines\parindent=0pt
      Amanullah R. et al., 2010, ApJ, 716, 712   

      Antoniou I, Perivolaropoulos L, 2010, JCAP, 12, 012

{     Atrio-Barandela F., Kashlinsky A., Ebeling H., Kocevski D., 2010, 
\quad Journal of Physics: Conference Series, 229, 012003}

      Clarkson, C. A., Maartens, R., 2010, Class. Quantum Grav., 27, 124008

      Dressler A., Faber S. M., 1990, ApJ, 354, 13

      Ehlers J., Sachs R., 1959, Z. Phys., 155, 498

      Ellis G. F. R., 2011, Class. Quantum Grav., 28, 164001 

      Ellis G. F. R., Tivon G., 1985, Observatory, 105, 189

      Feldman H. A., Watkins R., Hudson M. J., 2010, MNRAS, 407, 2328

      Freund J. E., Walpole R. E., 1980, Mathematical Statistics. Prentice-Hall, New Jersey,
\quad p.444

      Gurvits L. I., 1994, ApJ, 425, 442

      Hand N., et al., 2012, preprint(arXiv:1203.4219)

      Humphreys N. P., Maartens R., Matravers D. R., 1997, ApJ, 477, 47

      Jackson J. C., 2008, MNRAS, 390, L1

      Jackson J. C., Dodgson. M., 1996, MNRAS, 278, 603

      Jackson J. C., Dodgson. M., 1997, MNRAS, 285, 806

      Jackson J. C., 2004, JCAP, 11, 007

      Jackson J. C., Jannetta A.L., 2006, JCAP, 11, 002

      Kashlinsky A., Atrio-Barandela F., Kocevski D., Ebeling H., 2008, ApJ, 686, L49

      Kashlinsky A., Atrio-Barandela F., Kocevski D., Ebeling H., 2009, ApJ, 691, 1479

      Kashlinsky A., Atrio-Barandela F., Ebeling H., Edge A., Kocevski D., 2010,
\quad ApJ, 712, L81

      Keisler R., 2009, ApJ, 707, L42

      Kogut A., et al., 1993, ApJ, 419, 1 

      Kovalev Y. Y., et al., 2005, AJ, 130, 2473

{     Longo M. J., 2012, preprint(arXiv:1202.4433)}

      Lynden-Bell D., Faber S. M., Burstein D., Davies R. L., Dressler A., Terlevich R. J.,
\quad Wegner G., 1988, ApJ, 326, 19

      Maartens R., Humphreys N. P., Matravers D. R., Stoeger, W. R., 1996, 
\quad Class. Quantum Grav., 13, 253

{     Maartens R., Humphreys N. P., Matravers D. R., Stoeger, W. R., 1996, 
\quad Class. Quantum Grav., 13, 1689}

      Marscher A. P., Shaffer D. B., 1980, AJ, 85, 668

      Mather, J. C., Fixsen, D. J., Shafer, R. A., Mosier, C., Wilkinson, D. T., 1999,
\quad ApJ, 512, 511

      Mody K., Hajian A., 2012, preprint(arXiv:1202.1339)

      Osborne S. J., Mak D. S. Y., Church S. E., Pierpaoli E., 2011, ApJ, 737, 98

Paczynski B., Piran T., ApJ, 1990, 364, 341

{     Pashchenko I. N., Vitrishchak V. M., 2011, Azh, 2011, 88, 323}
     
      Pearson T. J., Readhead A. C. S., 1981, ApJ, 248, 61

      Penzias A. A., Wilson R.W., 1965, ApJ, 142, 419

      Percival W. J. et al., 2002, MNRAS, 337, 1068

      Pirani F. A. E., 1965, in Brandeis Lectures 1964 Vol. 1, General Relativity.
\quad Prentice-Hall, New Jersey, pp. 331-362

      Preston R. A., Morabito D. D., Williams J. G., Faulkner J., Jauncey D. L.,
\quad Nicolson G. D., 1985, AJ, 90, 1599

      Raine D. J., Thomas E. G., 1981, MNRAS, 195, 649

      Sachs R., 1961, Proc. R. Soc. Lond., A264, 309 

      Taylor G. B., Vermeulen R .C., Readhead A. C. S., Pearson T. J., Henstock D. R.,
\quad Wilkinson P. N., 1996, ApJS, 107, 37

      Thompson A. R., Moran J.M., Swenson G.W., Jr, 1986, Interferometry and Synthesis
\quad in Radio Astronomy, Wiley, New York, p.13

      Watkins R., Feldman H. A., Hudson M. J., 2009, MNRAS, 392, 743

      Wilson R. W., Penzias A.A., 1967, Sci, 156, 1100

{     Wright A., Otrupcek R., 1990, PKSCAT90, 1990 Parkes Catalogue Australia Telescope
\quad National Facility, http://www.parkes.atnf.csiro.au/research/surveys/pkscat90.html}

\end{document}